\documentclass[%
 reprint,
 amsmath,amssymb,
 aps,
 prd,
 nofootinbib,nobibnotes,
]{revtex4-2}

\usepackage{graphicx}
\usepackage{dcolumn}
\usepackage{bm}
\usepackage[dvipsnames, usenames]{xcolor}
\usepackage{acronym}
\usepackage[hidelinks]{hyperref}
\usepackage{physics}
\usepackage{orcidlink}

\graphicspath{{plots}}

\newcommand{\comment}[1]{}
\newcommand{\heading}[1]{\textbf{\textit{#1.}}}

\newcommand{\ligo}{\affiliation{LIGO Laboratory, Massachusetts Institute of Technology, Cambridge, MA 02139, USA}}
\renewcommand{\mit}{\affiliation{Kavli Institute for Astrophysics and Space Research and Department of Physics, Massachusetts Institute of Technology, Cambridge, MA 02139, USA}}

\begin{document}

\title{Nonparametric analysis of correlations in the binary black hole population with LIGO--Virgo--KAGRA data}

\author{Jack Heinzel\,\orcidlink{0000-0002-5794-821X}}
\email{heinzelj@mit.edu}
\ligo\mit

\author{Matthew Mould\,\orcidlink{0000-0001-5460-2910}}
\ligo\mit

\author{Salvatore Vitale\,\orcidlink{0000-0003-2700-0767}}
\ligo\mit

\date{March 17, 2025}

\begin{abstract}
Formation channels of merging compact binaries imprint themselves on the distributions and correlations of their source parameters, but current understanding of this population is hindered by simplified parametric models. We overcome such limitations using \textsc{PixelPop} [Heinzel \textit{et al.} (2025)]---our Bayesian nonparametric multidimensional population model. We analyze data from the first three LIGO--Virgo--KAGRA observing runs and make high resolution, minimally modeled measurements of the pairwise distributions of binary black hole masses, redshifts, and spins. We find no evidence that the mass spectrum evolves over redshift and show that such measurements are fundamentally limited by the detector horizon. We find support for correlations of the spin distribution with binary mass ratio and redshift, but at reduced significance compared to overly constraining parametric models. Confident data-driven conclusions about population-level correlations with flexible models like \textsc{PixelPop} will require more informative gravitational-wave catalogs.
\end{abstract}

\maketitle

\heading{Introduction}---From the first gravitational-wave (GW) detection of a merging black hole (BH) binary \cite{TheLIGOScientific:2016wfe} to the catalog observed by the end of the third observing run (O3) \cite{LIGOScientific:2018mvr, LIGOScientific:2020ibl, LIGOScientific:2021usb, LIGOScientific:2021djp} of the LIGO--Virgo--KAGRA collaboration (LVK) \cite{LIGOScientific:2014pky, VIRGO:2014yos, KAGRA:2020tym}, our understanding of the underlying population has improved significantly \cite{LIGOScientific:2018jsj, LIGOScientific:2020kqk, KAGRA:2021duu}: the merger rate increases with redshift, consistent with cosmic star-formation history \cite{Madau:2014bja, Fishbach:2018edt, Callister:2020arv, Schiebelbein-Zwack:2024roj}; there is a continuum of masses with peaks at around $10M_\odot$ and $35M_\odot$, with most binaries having nearly equal masses \cite{Talbot:2018cva, Tiwari:2021yvr, Farah:2023vsc}; and there are relatively fewer BHs with large spins \cite{Callister:2022qwb, Mould:2022xeu, Tong:2022iws, Fishbach:2021xqi}. Potential astrophysical correlations between source parameters have also been identified \cite{Callister:2021fpo, Biscoveanu:2022qac, Franciolini:2022iaa, Li:2023yyt, Pierra:2024fbl}. Despite this, the physical origins of these features are unclear, though there are a plethora of potential formation channels (e.g., Refs. \cite{Ivanova:2012vx, vandenHeuvel:2017pwp, Gallegos-Garcia:2021hti, Mandel:2015qlu, deMink:2016vkw, Marchant:2016wow, Downing:2009ag, Rodriguez:2015oxa, Rodriguez:2019huv, Mapelli:2021gyv, Miller:2008yw, Antonini:2016gqe, Mckernan:2017ssq, Stone:2016wzz, Silsbee:2016djf, Liu:2018nrf}) with distinct GW signatures (e.g., Refs \cite{Bavera:2020uch, Bavera:2022mef, Zevin:2020gbd, Zevin:2022wrw, Broekgaarden:2022nst, Fuller:2019sxi, Bavera:2020inc, Fuller:2022ysb, Barkat:1967zz, Woosley:2016hmi, Tanikawa:2021zfm}). Interpreting GW data in light of these predictions could yield key astrophysical insights.

The above population constraints are made using a model for the merger rate density as a function of the binary source parameters, for which the combined catalog of GW events can be used to infer a Bayesian posterior \cite{Mandel:2018mve, Thrane:2018qnx, Vitale2020, Essick:2023upv}. Due to large theoretical uncertainties \cite{Mandel:2021smh}, a major difficulty is choosing appropriate models. Perhaps the most common approach due to its simplicity is to parametrize the rate density with basic functions (such as power laws and normal distributions) and infer their parameters (e.g., Refs. \cite{KAGRA:2021duu, Fishbach:2018edt, Talbot:2018cva, Wysocki:2018mpo, Talbot:2017yur, Miller:2020zox, Roulet:2018jbe}). However, such choices can be overly constraining and lead to missed features in the inferred population or, conversely, features driven by the strong model assumptions rather than data \cite{Vitale:2022dpa, Farah:2021qom, Farah:2023swu}. An alternative approach is to use nonparametric models in order to impose any assumptions as weakly as possible \cite{Mandel:2016prl, KAGRA:2021duu, Edelman:2022ydv, Golomb:2022bon, Payne:2022xan, Heinzel:2023hlb, Rinaldi:2021bhm, Callister:2023tgi, Ray:2023upk, Toubiana:2023egi}. Such flexible models are better able to fit complicated structures in the source distributions that may arise from astrophysical formation processes, though typically at the cost of larger uncertainties, higher computational cost, and lack of interpretability. Moreover, it is challenging to extend these methods to simultaneously probe multiple parameter dimensions and their correlations with high fidelity \cite{KAGRA:2021duu, Sadiq:2023zee, Ray:2023upk, Ray:2024hos, Fishbach:2019bbm, Farah:2023swu}.

In Ref. \cite{Heinzel:2024jlc} we developed \textsc{PixelPop}---a Bayesian nonparametric multidimensional population model for inference on correlated parameter distributions, such as those that may result from the astrophysical formation of compact binaries. \textsc{PixelPop} works by densely binning the joint space of binary source parameters and inferring the comoving merger rate density in each bin. The only assumption made is a weak smoothing prior that couples each bin to its nearest neighbors---the conditional autoregressive (CAR) prior. The computational efficiency of the CAR model allows us to dramatically increase its resolution compared to similar nonparametric methods \cite{Mandel:2016prl, KAGRA:2021duu, Ray:2023upk}---using $\mathcal{O}(10^4)$ bins compared to $\mathcal{O}(10^2)$---and thus provide a much more complete view of GW populations while making fewer assumptions about them.

In this Letter, we analyze the LVK catalog of binary BH mergers with \textsc{PixelPop}, including the 69 binary BH events with false-alarm rates $<1\,\mathrm{yr}^{-1}$ in the third GW transient catalog (GWTC-3) \cite{LIGOScientific:2021djp, KAGRA:2021duu}. We specifically target the joint distributions of: (1) the heavier (primary) BH mass $m_1$ and redshift $z$, (2) binary mass ratio $0<q\leq1$ and effective aligned spin $\chi_\mathrm{eff}\in(-1,1)$ \cite{Racine:2008qv}, and (3) $\chi_\mathrm{eff}$ and $z$ (when not inferred with \textsc{PixelPop} we fit the mass, redshift, and spin distributions with the parametric \textsc{Power Law + Peak} mass model, \textsc{Power Law} redshift model, and \textsc{Default} spin model from Ref.~\cite{KAGRA:2021duu}). We use the publicly available \cite{LIGOScientific:2019lzm, KAGRA:2023pio} parameter-estimation samples \cite{gwtc1_pe, gwtc2_pe, ligo_scientific_collaboration_and_virgo_2022_6513631, ligo_scientific_collaboration_and_virgo_2023_8177023} (\texttt{Overall}, \texttt{PrecessingSpinIMRHM}, and \texttt{Mixed} for events from GWTC-1, 2, and 3, respectively) and sensitivity estimates \cite{ligo_scientific_collaboration_and_virgo_2023_7890398} to compute the GW population likelihood, following a Monte Carlo approximation \cite{Thrane:2018qnx, Vitale2020, KAGRA:2021duu} and penalizing the likelihood in regions of high approximation uncertainty \cite{Farr:2019rap, Essick:2022ojx, Talbot:2023pex}. Full details of \textsc{PixelPop} and its analysis methods are available in \citeauthor{Heinzel:2024jlc} \cite{Heinzel:2024jlc}. We perform analyses on additional parameter pairs in the Supplemental Material.

\begin{figure}
\centering
\includegraphics[width=1\linewidth]{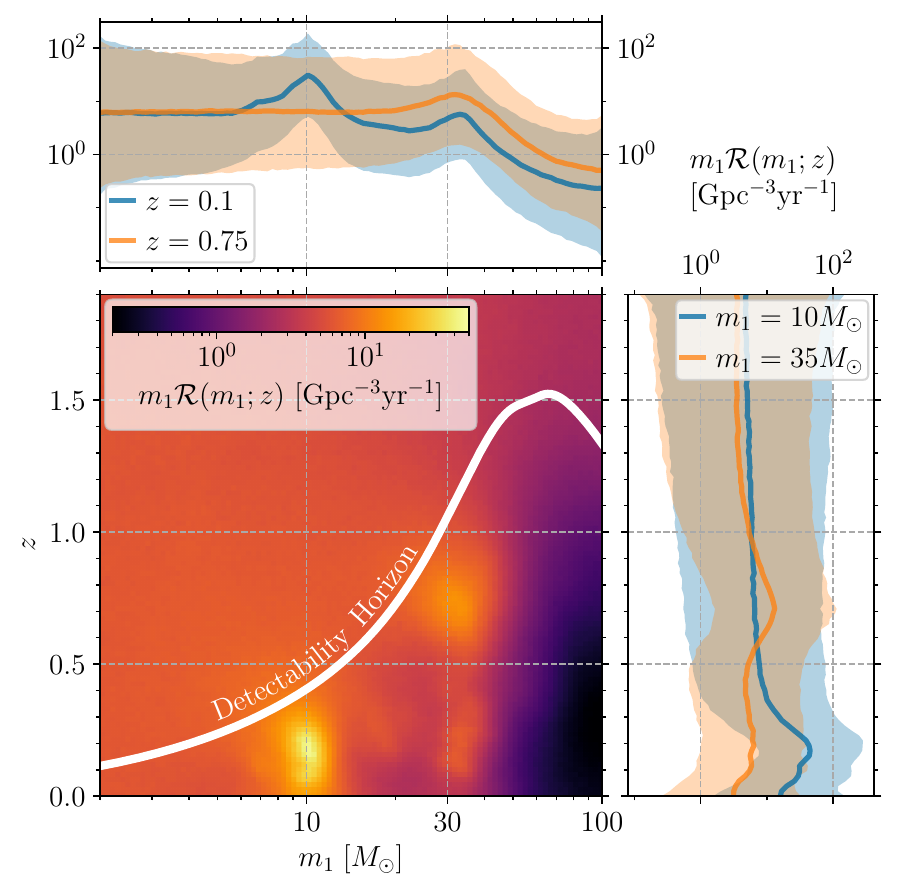}
\caption{Comoving merger rate density $\mathcal{R}(\ln m_1;z) = m_1 \mathcal{R}(m_1;z)$ as a function of primary mass $m_1$ and redshift $z$, inferred from GWTC-3. The central panel shows the posterior median, with lower to higher rates colored black to yellow, respectively. The white curve defines the detection horizon above which a binary BH merger is not detectable. The upper horizontal panel shows $\mathcal{R}$ as a function of $m_1$ at two fixed redshifts, $z=0.1$ (blue) and $z=0.75$ (orange). The solid line is the posterior median while the shaded region encloses the 90\% symmetric credible region. Similarly, the right-hand vertical panel shows $\mathcal{R}$ as a function of $z$ at the two peaks in the merger rate of $m_1=10M_\odot$ (blue) and $m_1=35M_\odot$ (orange).}
\label{fig:m1z}
\end{figure}

\heading{Primary mass and redshift}---The observed primary mass distribution spans roughly two orders of magnitude, so we place 100 bins uniformly over both $\ln m_1$ from $m_1=2M_\odot$ to $m_1=100M_\odot$ and redshift from $z=0$ to $z=1.9$ ($10^4$ bins total). In Fig. \ref{fig:m1z} we show the comoving merger rate density $\mathcal{R}(m_1;z)$ inferred from GWTC-3 over $m_1$ and $z$; note that $\mathcal{R}(m_1;z)$ is a function of redshift $z$ through the comoving volume element, but is not a density in $z$. A representative average over the posterior uncertainty is given by the median rate bins in the central panel. One dimensional slices of the joint distribution and their associated uncertainties are displayed in the upper and right-hand panels.

We observe peaks at $m_1 \approx 10 M_\odot$ and $m_1 \approx 35 M_\odot$,
consistent with previous parametric and nonparametric results \cite{Talbot:2018cva, KAGRA:2021duu, Edelman:2021zkw, Edelman:2022ydv, Godfrey:2023oxb, Farah:2023vsc, Callister:2023tgi}. For $z \in [0,0.2]$, the merger rate density integrated over $m_1 / M_\odot \in [8,15]$ is higher than when integrated over $m_1 / M_\odot \in [8,100]$ at almost 100\% posterior credibility. The secondary peak has somewhat lower significance in the same redshift interval; we quantify this with the local integrated rate density over $m_1 / M_\odot \in [25,40]$ and $m_1 / M_\odot \in [15,50]$, where the former is higher than the latter at 72\% credibility.

Within the posterior uncertainty we find no evidence for structure beyond these two peaks. But unlike for parametric mass models, the inferred merger rate does not decrease at small masses and the uncertainty grows as there is insufficient information in the catalog to constrain the merger rate with the flexibility of \textsc{PixelPop}. This is to be expected because at low masses the detector horizon---the redshift at which a source is not detectable, given by the white curve\footnote{We compute this as the maximum redshift at which a source has optimal signal-to-noise ratio $>8$ in a LIGO--Virgo detector network with O3-like sensitivity \cite{KAGRA:2013rdx}.} in Fig. \ref{fig:m1z}---excludes most of the mass range. In the absence of GW events, inference in this region of parameter space is dominated by the CAR prior, similar to other nonparametric models \cite{Callister:2023tgi, Farah:2024xub}. In contrast, the merger rate decreases at large masses and low redshifts where sources should be detectable; having not detected such sources implies a low astrophysical merger rate.

There is a mild preference for an increasing merger rate at small redshifts $z \lesssim 0.2$, consistent with parametric models \cite{Fishbach:2018edt, Callister:2020arv, KAGRA:2021duu, Schiebelbein-Zwack:2024roj}. For masses $m_1 / M_\odot \in [8,100]$ beyond the prior-dominated regime, the integrated comoving rate density is higher over $z \in [0.1,0.2]$ than $z \in [0,0.1]$ at 67\% credibility. At $m_1 = 10M_\odot$, the comoving merger rate increases from $z=0$ to $z=0.2$ at 73\% credibility. With monotonic parametric models \cite{KAGRA:2021duu, Fishbach:2018edt}, a positive slope at low redshifts enforces the same at higher redshifts, whereas \textsc{PixelPop} does not make this requirement. There appears to be another redshift mode at $z\approx0.7$ for masses within the detection horizon, coincident with the onset of a plateau in the merger rate inferred in Ref.~\cite{Callister:2023tgi}. At $m_1 = 35M_\odot$, the comoving merger rate increases from $z=0$ to $z=0.7$ at 79\% credibility. Above this redshift the posterior uncertainty increases as the majority of the parameter space lies beyond the detector horizon where the CAR model prefers broad distributions.

Beyond this, we do not confidently identify any correlations in the mass--redshift parameter space, in agreement with Ref.~\cite{Ray:2023upk,Sadiq:2021fin}. While the $m_1 \approx 35 M_\odot$ peak extends out to the $z \approx 0.7$ mode and beyond, we cannot conclude that the mass spectrum evolves over redshift as suggested in Ref.~\cite{Rinaldi:2023bbd} because the $m_1 \approx 10 M_\odot$ mode is limited by the detector horizon. We cannot exclude the possibility that the lower-mass peak has the same redshift extent as the higher-mass peak.

\heading{Mass ratio and effective spin}---The effective spin $\chi_\mathrm{eff}$ is the mass-weighted average of the two BH spin components along the binary orbital angular momentum; it is large and positive (negative) for binaries with spinning BHs aligned (antialigned) with the orbital angular momentum, and closer to zero for small or misaligned BH spins. \citeauthor{Callister:2021fpo}~\cite{Callister:2021fpo} originally discovered evidence for an anticorrelation between mass ratios $q$ and effective spins $\chi_\mathrm{eff}$---i.e., binaries with larger $\chi_\mathrm{eff}$ tend to have more unequal masses---with a strongly parameterized model. This result was corroborated by Refs.~\cite{Adamcewicz:2022hce, Adamcewicz:2023mov, Heinzel:2023hlb, KAGRA:2021duu} with additional models, though with reduced confidence in regions of uninformative data \cite{Heinzel:2023hlb}. However, mechanisms that could produce this correlation are not well understood \cite{Broekgaarden:2022nst, Bavera:2020uch, Zevin:2022wrw, Gerosa:2021mno, McKernan:2021nwk, Santini:2023ukl, Baibhav:2022qxm, Santini:2023ukl, Olejak:2021iux, Olejak:2024qxr} and the models used make strong assumptions about the form of the correlation. Instead, we quantify the evidence for this correlation and constrain its form with minimal assumptions by fitting the merger rate density $\mathcal{R}(\chi_\mathrm{eff},q;z)$ with 100 bins over both $\chi_\mathrm{eff}$ and $q$ using \textsc{PixelPop}.

\begin{figure}
\centering
\includegraphics[width=1\linewidth]{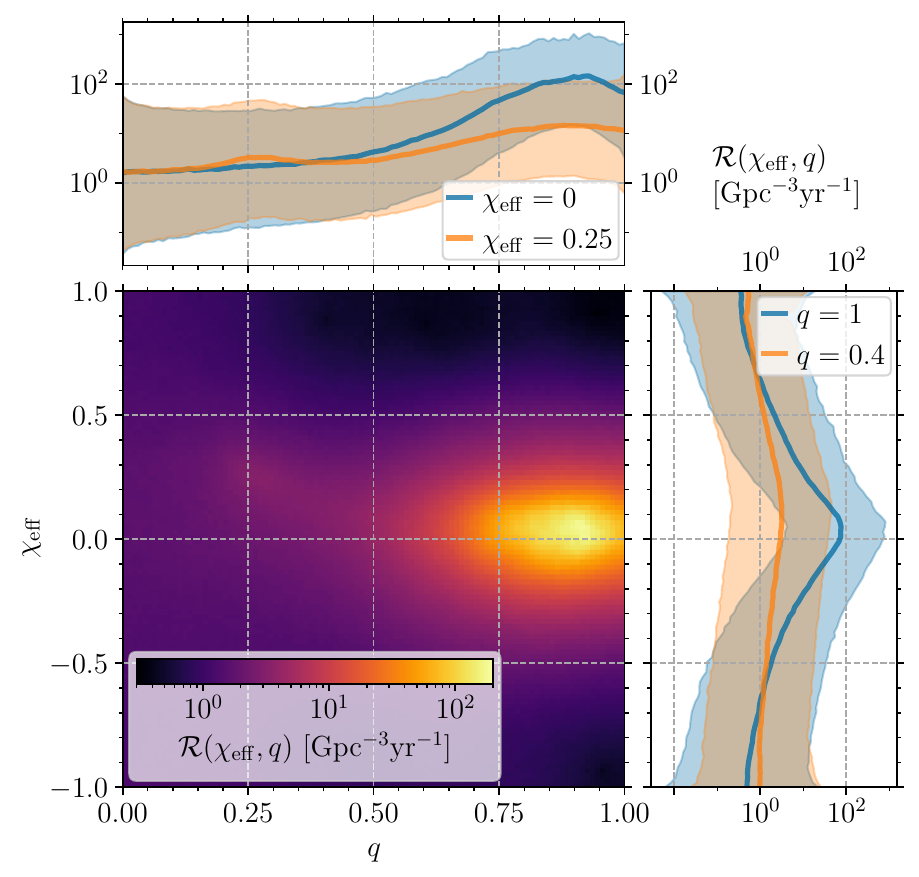}
\caption{Joint comoving merger rate density $\mathcal{R}(\chi_\mathrm{eff},q;z=0.2)$ as a function of effective spin $\chi_\mathrm{eff}$ and binary mass ratio $q$, evaluated at a fixed redshift $z=0.2$. The central panel displays the joint two-dimensional posterior median. The merger rate density evaluated at $\chi_\mathrm{eff}=0$ (blue) and $\chi_\mathrm{eff}=0.25$ (orange) is shown as a function of $q$ in the upper panel, and evaluated at $q=1$ (blue) and $q=0.4$ (orange) as a function of $\chi_\mathrm{eff}$ in the right-hand panel. For the one-dimensional slices the bold line gives the posterior median while the shaded region encloses the 90\% symmetric credible region.}
\label{fig:qchi_median}
\end{figure}

In Fig. \ref{fig:qchi_median}, we show the inferred comoving merger rate density evaluated at a redshift of $z=0.2$, corresponding to the peak in Fig. \ref{fig:m1z}. The merger rate is highest for near-equal masses and $\chi_\mathrm{eff} \approx 0$ and decreases towards unequal masses and $|\chi_\mathrm{eff}|>0$, though the data do not require the global maximum to be $q=1$ and $\chi_\mathrm{eff}=0$. \textsc{PixelPop} does not enforce that the merger rate vanishes as $q \to 0$ (unlike common parametric models \cite{KAGRA:2021duu}) and there is also support for negative $\chi_\mathrm{eff}$, though the tails of the distributions again become dominated by the CAR prior (cf. Ref. \cite{Callister:2023tgi}). The bulk of the population has $q>0.6$ and in this region the integrated merger rate is higher for $\chi_\mathrm{eff} \in [0,0.5]$ than $\chi_\mathrm{eff} \in [-0.5,0]$ at 91.7\% posterior credibility, indicating that component BH spins tend to prefer alignment with the orbital angular momentum \cite{KAGRA:2021duu}. Specifically, the merger rate density at $z=0.2$ is $11.3_{-3.8}^{+5.8} \; \mathrm{Gpc}^{-3} \mathrm{yr}^{-1}$ for positive spins $\chi_\mathrm{eff} \in [0,0.5]$ (median and 90\% credible interval) but $7.4_{-3.1}^{+4.8} \; \mathrm{Gpc}^{-3} \mathrm{yr}^{-1}$ for negative $\chi_\mathrm{eff} \in [-0.5,0]$, in both cases taking $q \in [0.6, 1]$.

The two-dimensional posterior median in Fig.~\ref{fig:qchi_median} suggests a tail in the distribution toward $(q, \chi_\mathrm{eff}) \approx (0.25, 0.25)$. In the region $q \in [0.2, 0.4]$, we find a higher rate density for $\chi_\mathrm{eff} \in [0, 0.5]$ than $\chi_\mathrm{eff} \in [-0.5,0]$ at 66\% credibility when accounting for the posterior uncertainty. To further test for the aforementioned $q$--$\chi_\mathrm{eff}$ correlation, we compute the Spearman rank correlation coefficient $\rho$ \cite{Spearman:1904abc,kendall1979advanced}---which is positive (negative) when there is an increasing (decreasing) monotonic correlation---for $\mathcal{R}(\chi_\mathrm{eff},q)$, restricting to $q \in [0.2, 1]$ and $\chi_\mathrm{eff} \in [-0.5,0.5]$ to avoid prior-dominated regions. Whereas Ref.~\cite{KAGRA:2021duu} identify a negative correlation between $q$ and $\chi_\mathrm{eff}$ at 97.5\% significance, we find $\rho = 0.023_{-0.270}^{+0.281}$ and no significant evidence for a correlation. We stress, however, that the \textsc{PixelPop} posterior does not exclude such a correlation and that this result is not at odds with the results of strongly parametrized models \cite{Callister:2021fpo, KAGRA:2021duu, Adamcewicz:2022hce, Heinzel:2023hlb} when used to analyze the same GW catalog---if one mandates a functional form that allows a global monotonic trend then the data prefer an anticorrelation, but this is neither required nor excluded by the more flexible \textsc{PixelPop} model that allows a much broader class of nontrivial correlations.

\heading{Redshift and effective spin}---\citeauthor{Biscoveanu:2022qac} \cite{Biscoveanu:2022qac} found evidence for an astrophysical broadening of the $\chi_\mathrm{eff}$ distribution with increasing redshift, assuming a linear relationship. This was confirmed by \citeauthor{Heinzel:2023hlb} \cite{Heinzel:2023hlb} with a more flexible model for the correlation, but in both cases a simple Gaussian parametrization was made for the underlying population. We remove these assumptions by using \textsc{PixelPop} to infer the comoving merger rate density $\mathcal{R}(\chi_\mathrm{eff};z)$ jointly over 100 bins in each of $\chi_\mathrm{eff}$ and $z$.

\begin{figure}
\centering
\includegraphics[width=1\linewidth]{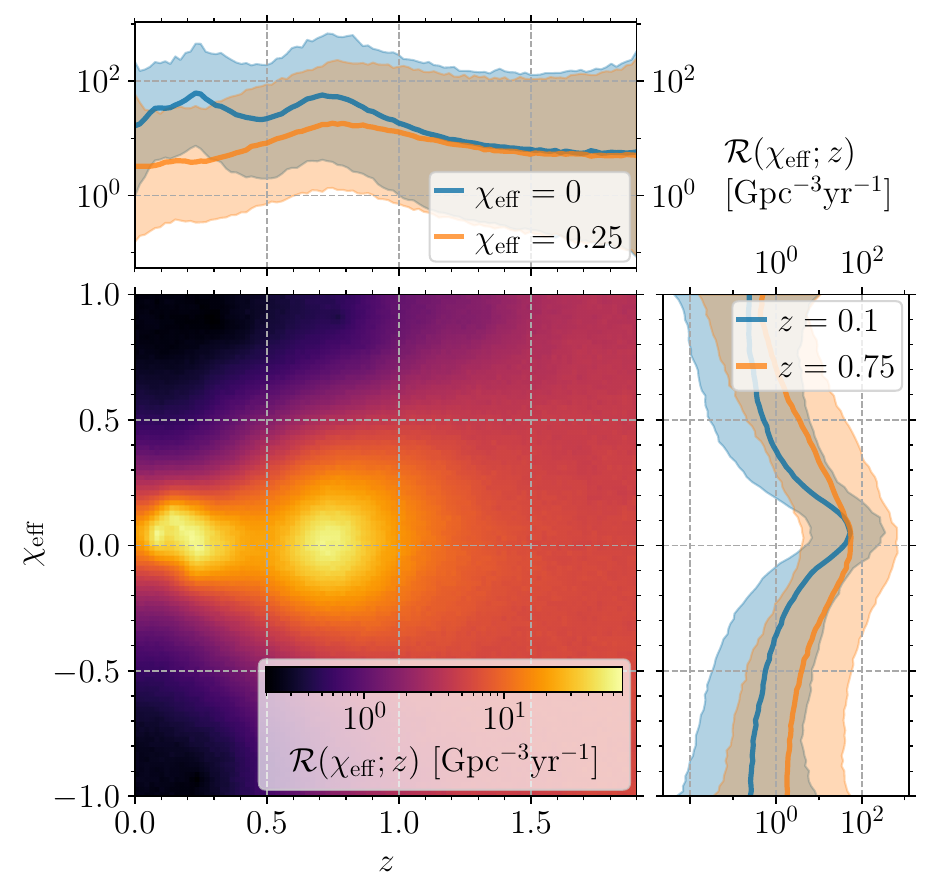}
\caption{
Comoving merger rate density $\mathcal{R}(\chi_\mathrm{eff};z)$ as a function of effective spin $\chi_\mathrm{eff}$ and redshift $z$. The central panel displays the joint two-dimensional posterior median. The merger rate density evaluated at $\chi_\mathrm{eff}=0$ (blue) and $\chi_\mathrm{eff}=0.25$ (orange) is shown as a function of $z$ in the upper panel, and evaluated at $z=0.1$ (blue) and $z=0.75$ (orange) as a function of $\chi_\mathrm{eff}$ in the right-hand panel. For the one-dimensional slices the bold line gives the posterior median while the shaded region encloses the 90\% symmetric credible region.
}
\label{fig:zchi_median}
\end{figure}

We show the \textsc{PixelPop} posterior in Fig.~\ref{fig:zchi_median}. The merger rate may evolve differently over redshift depending on the effective spin: for $\chi_\mathrm{eff}=0$ the median rate density is initially higher at lower redshifts than $\chi_\mathrm{eff}=0.25$, increases from $z=0$ to $z\approx0.25$, then drops at $z\approx0.5$ before increasing again at $z\approx0.7$; for $\chi_\mathrm{eff}=0.25$ the median rate density increases monotonically between $z=0$ and $z\approx0.7$; in both cases the rate density plateaus in the prior-dominated region $z\gtrsim1$. While these results hold for the median inferred rate, this marginalizes over large uncertainties, within which a lack of redshift peaks is also possible.

We can confidently conclude that the merger rate significantly decreases for $|\chi_\mathrm{eff}|\gtrsim0.5$ at low redshifts where sources would be detectable but have not been detected, implying that the astrophysical merger rate must be low. This corresponds to the broadening of the $\chi_\mathrm{eff}$ spin distribution visible in both the two-dimensional median posterior rate density and the one-dimensional slices in Fig.~\ref{fig:zchi_median}. There is a more pronounced peak just above $\chi_\mathrm{eff}\approx0$ at $z=0.1$ compared to $z=0.75$ where the overall rate is higher. We quantify this trend by computing the Spearman rank correlation coefficient $\rho$ between the redshift and the width of the effective spin distribution---if $\rho$ is positive, the $\chi_\mathrm{eff}$ distribution broadens at larger redshifts. However, we must be careful not to mistake the natural broadening of the CAR model in uninformative regions of parameter space---such as higher redshifts---for an astrophysical correlation. For $z<0.5$, we find a weak preference for broadening, with $\rho = 0.096_{-0.240}^{+0.231}$ and $\rho > 0$ at 75\% credibility; cf. 98\% in Refs.~\cite{Biscoveanu:2022qac,Heinzel:2023hlb}. Limiting to $z<1$ instead yields $\rho = 0.164_{-0.236}^{+0.267}$ and $\rho>0$ at a larger but still mild significance of 84\%. The broadening of $\chi_\mathrm{eff}$ over redshift appears to be a genuine astrophysical feature in the data, but with \textsc{PixelPop} we cannot make a definitive statement about the underlying nature of the population without more informative data.

\heading{Conclusions}---As the catalog of binary BH mergers observed with GW detectors grows, increasingly more information can be obtained about the underlying population. In particular, nontrivial structures and population-level parameter correlations may be identified and interpreted in terms of their astrophysical origins and formation channels, but only by using a suitable population model. Indeed, correlations have been identified in the population of binary BHs with simplified parametric models \cite{Callister:2021fpo, Biscoveanu:2022qac, Heinzel:2023hlb, Adamcewicz:2022hce, Adamcewicz:2023mov, Franciolini:2022iaa, Pierra:2024fbl}, but these risk misspecification and overconfident inference.

\textsc{PixelPop} \cite{Heinzel:2024jlc} is a nonparametric multidimensional population model that imposes minimal assumptions on the form of the underlying merger rate and is computationally efficient. We used \textsc{PixelPop} to analyze bivariate source distributions with high resolution---primary BH masses and redshifts, binary mass ratios and effective spins, and effective spins and redshifts---using LVK binary BH detections \cite{LIGOScientific:2018mvr, LIGOScientific:2020ibl, LIGOScientific:2021usb, LIGOScientific:2021djp}. We showed that: there is no evidence for an evolving BH mass spectrum over redshift and this measurement is currently prohibited by the detector horizon; and the GW data are consistent with a distribution of effective spins that is correlated with binary mass ratio and redshift, but that we cannot be confident in this hypothesis without strong model assumptions.

Flexible population inference methods such as Bayesian nonparametrics will be increasingly useful in the future as GW catalogs become more informative. They offer a valuable consistency check that strongly parameterized models are not overfitting to or extrapolating from observational data, as the shapes of distributions as well as correlations in their joint space may be complicated and impossible to model with simple parameterizations. Flexible models can also be used to account for unmodeled contributions to the GW population that would otherwise lead to systematically biased inference \cite{Cheng:2023ddt}, or for cosmological constrains with astrophysics-agnostic assumptions about the source population \cite{MaganaHernandez:2024uty, Farah:2024xub}. The ability of \textsc{PixelPop} to capture arbitrary correlations between source parameters make it a standout model for such GW population analyses.

\heading{Acknowledgements}---We thank Sofía Álvarez-López, Jacob Golomb, Cailin Plunkett, Noah Wolfe, and the Rates and Populations LIGO working group for useful discussions and helpful comments.
J.H is supported by the NSF Graduate Research Fellowship under Grant No. DGE1122374.
M.M. is supported by the LIGO Laboratory through the National Science Foundation award PHY-1764464.
J.H. and S.V. are partially supported by the NSF grant PHY-2045740.
This material is based upon work supported by NSF's LIGO Laboratory which is a major facility fully funded by the National Science Foundation.
The authors are grateful for computational resources provided by the LIGO Laboratory and supported by National Science Foundation Grants PHY-0757058 and PHY-0823459.

\bibliography{main}

\clearpage

\section{Supplemental Material}

We include the analysis of additional pairwise correlations in LIGO--Virgo--KAGRA (LVK) data from the first three gravitational-wave (GW) catalogs using \textsc{PixelPop}. Below, we infer the joint distributions of: (1) primary binary black hole (BH) mass $m_1$ and effective spin $\chi_\mathrm{eff}$; (2) $m_1$ and binary mass ratio $q$; and (3) BH spin magnitudes $a$ and orbital misalignment angles $\theta$.

\begin{figure}
\centering
\includegraphics[width=1\linewidth]{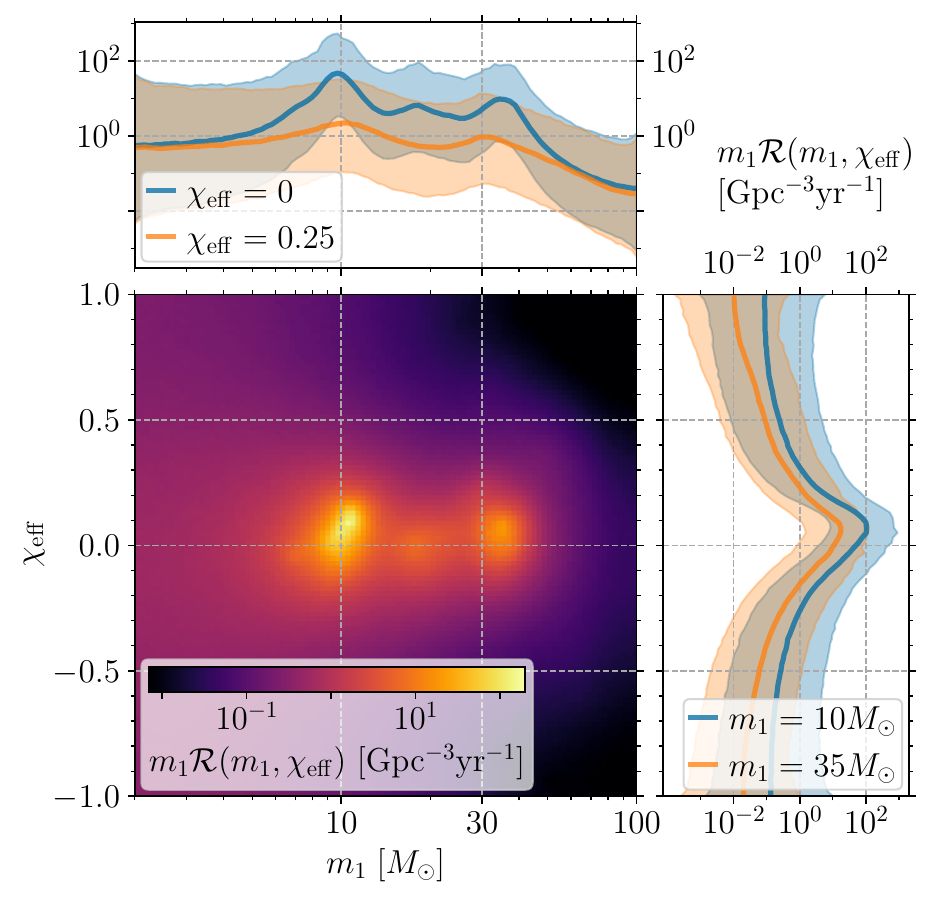}
\caption{
Comoving merger rate density $\mathcal{R}(\ln m_1, \chi_{\rm eff}; z) = m_1\mathcal{R}(m_1, \chi_{\rm eff}; z)$ as a function of primary mass $m_1$ and effective spin $\chi_\mathrm{eff}$, evaluated at a fixed redshift $z=0.2$. The central panel displays the joint two-dimensional posterior median. The merger rate density evaluated at $\chi_\mathrm{eff}=0$ (blue) and $\chi_\mathrm{eff}=0.25$ (orange) is shown as a function of $m_1$ in the upper panel, and evaluated at $m_1=10 M_\odot$ (blue) and $m_1=35 M_\odot$ (orange) as a function of $\chi_\mathrm{eff}$ in the right-hand panel. For the one-dimensional slices the bold line gives the posterior median while the shaded region encloses the 90\% symmetric credible region.}
\label{fig:m1chi}
\end{figure}

\heading{Primary mass and effective spin}---\citeauthor{Ray:2024hos} \cite{Ray:2024hos} find evidence that the $\chi_\mathrm{eff}$ distribution is different for merging binary BHs with masses in the range $[30M_\odot,40M_\odot]$, using a three-dimensional binned Gaussian process prior to model primary mass, secondary mass, and effective spin. We use \textsc{PixelPop} to infer the merger rate density $\mathcal{R}(m_1,\chi_\mathrm{eff};z)$ with 100 uniform bins in both $\ln m_1$ between $m_1=2M_\odot$ and $m_1=100M_\odot$ and $\chi_\mathrm{eff}$ over $(-1,1)$. We plot the inferred merger-rate density evaluated at $z=0.2$ in Fig.~\ref{fig:m1chi}.

We find no evidence for different effective-spin distributions at different primary masses. Due to the different nonparametric approaches, this is not necessarily at odds with the findings of \citeauthor{Ray:2024hos} \cite{Ray:2024hos}. Since we do not jointly model the secondary masses in our analysis, this may suggest that when modeled independently, the two component BH mass distributions \cite{Fishbach:2019bbm, Farah:2023swu} have different $\chi_\mathrm{eff}$ distributions. However, no evidence for this has previously been found with parametric population models \cite{Mould:2022xeu}. Interestingly, when modeling the joint $m_1$--$\chi_\mathrm{eff}$ distribution an additional mode between $10<m_1/M_\odot<30$ appears at $\chi_\mathrm{eff}=0$, but at low significance.

\begin{figure}
\centering
\includegraphics[width=1\linewidth]{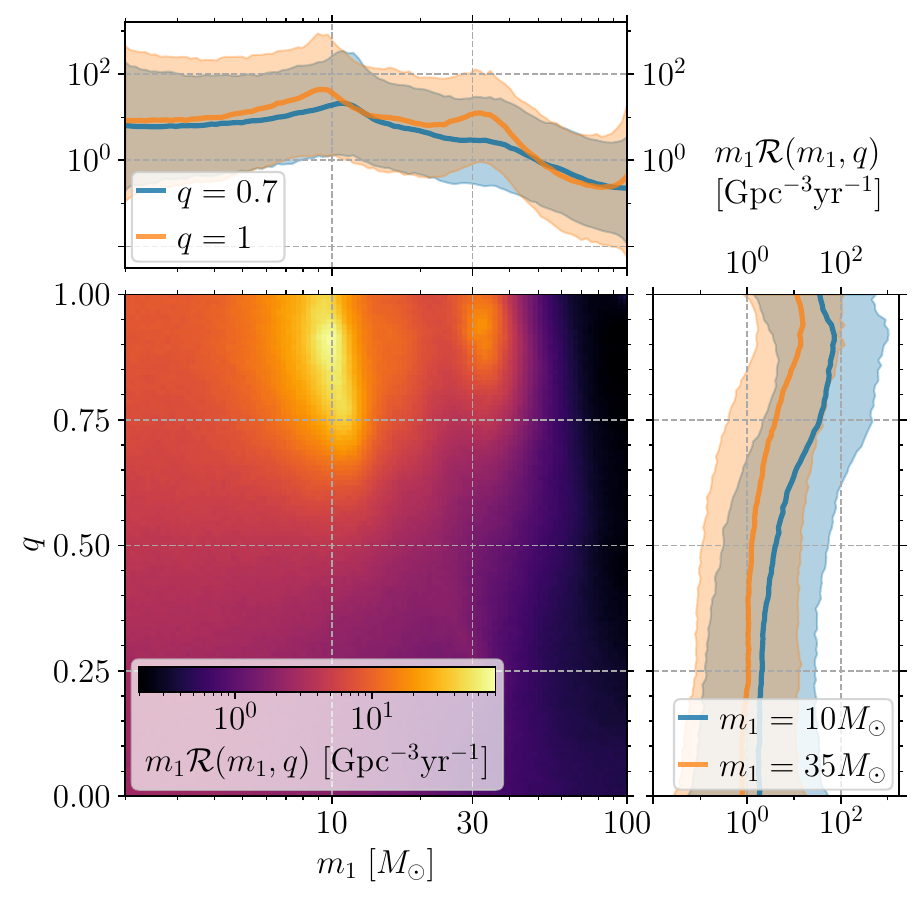}
\caption{
Comoving merger rate density $\mathcal{R}(\ln m_1,q;z) = m_1\mathcal{R}(m_1,q;z)$ as a function of primary mass $m_1$ and mass ratio $q$, evaluated at a fixed redshift $z=0.2$. The central panel displays the joint two-dimensional posterior median. The merger rate density evaluated at $q=0.7$ (blue) and $q=1$ (orange) is shown as a function of $m_1$ in the upper panel, and evaluated at $m_1=10 M_\odot$ (blue) and $m_1=35 M_\odot$ (orange) as a function of $q$ in the right-hand panel. For the one-dimensional slices the bold line gives the posterior median while the shaded region encloses the 90\% symmetric credible region.}
\label{fig: m1q}
\end{figure}

\heading{Primary mass and mass ratio}---To assess differences between the component mass distributions, we run \textsc{PixelPop} to infer the merger-rate density $\mathcal{R}(m_1,q;z)$ jointly over primary mass $m_1$ and mass ratio $q$ with the same bin placement as previously. As \textsc{PixelPop} allows for any correlations over the parameter space, this is entirely equivalent to modeling the joint distribution of component masses with a reparametrization, as in Refs. \cite{KAGRA:2021duu, Sadiq:2023zee, Ray:2023upk, Ray:2024hos, Fishbach:2019bbm, Farah:2023swu}. We show the \textsc{PixelPop} results, evaluated at $z=0.2$, in Fig.~\ref{fig: m1q}.

There is a potential trend in the primary mass distribution shifting to larger masses for mass ratios further from unity, and the marginal mass-ratio distribution may be flatter at larger primary masses. However, within the large \textsc{PixelPop} posterior uncertainties the merger rate may also be uncorrelated and even flat over component masses.

\heading{Component spins}---We also perform a nonparametric inference on the joint distribution of the component spins using \textsc{PixelPop}. These are specified by the dimensionless spin magnitudes $a_1$, $a_2$ and the polar angles $\theta_1$, $\theta_2$ between the component BH spin vectors and the binary orbital angular momentum for the primary (1) and secondary (2) BHs, respectively. We assume that the spins are independent and identically (IID) distributed. Phrased in terms of the comoving differential merger rate density $\mathcal{R} ( a , \cos\theta ; z )$, this means that
\begin{align}
& \mathcal{R} ( a_1 , \cos\theta_1 , a_2 , \cos\theta_2 ; z )
\nonumber \\
& =
\frac
{ \mathcal{R} ( a_1 , \cos\theta_1 ; z ) \mathcal{R} ( a_2 , \cos\theta_2 ; z ) }
{ \int \dd{a} \dd{(\cos\theta)} \mathcal{R} ( a , \cos\theta ; z ) }
\, .
\end{align}
This enforces the IID probability density for the component BH spins without double counting the overall merger rate.

As the merger rate model contributes twice due to the IID assumption, the uncertainty in the Monte Carlo likelihood estimator mentioned in the Introduction and described in detail in Ref.~\cite{Heinzel:2024jlc} is effectively doubled. We therefore replace the parameter-estimation samples for a subset of GW events (GW150914, GW170608, GW170729, GW170818, GW170823, GW200129\_065458, and GW200112\_155838) with the samples from the \textsc{IMRPhenomXPHM} \cite{Pratten:2020ceb} analyses. This increases the minimum sample count over all events from 3,194 to 14,802, thereby reducing Monte Carlo uncertainty in estimating the likelihood for each event. Additionally, we create a custom set of $\approx 10^9$ simulated GW signals using the \textsc{IMRPhenomXP} waveform approximant \cite{Pratten:2020ceb}, added to Gaussian noise colored by representative power spectral densities from the first, second, and third LVK observing runs, selected according to their respective observing durations \cite{KAGRA:2013rdx}. Sources are drawn from a fixed population using the \textsc{Power Law + Peak} mass model and \textsc{Power Law} redshift model, with parameters chosen to match the distributions inferred in Ref.~\cite{KAGRA:2021duu}. The spin magnitudes are drawn uniformly over $[0,1)$ and their directions are drawn isotropically. The remaining parameters are drawn from the default uniform parameter-estimation priors \cite{LIGOScientific:2021djp}. We approximate the detection criterion by selecting signals with a network matched-filter signal-to-noise ratio $>9$ \cite{Essick:2023toz}, which leaves $\approx1.4\times10^6$ found signals, compared to $\approx4\times10^4$ in the public sensitvity estimates \cite{ligo_scientific_collaboration_and_virgo_2023_7890398}, thereby significantly reducing the Monte Carlo uncertainty when estimating selection effects.

\begin{figure}
\centering
\includegraphics[width=0.95\linewidth]
{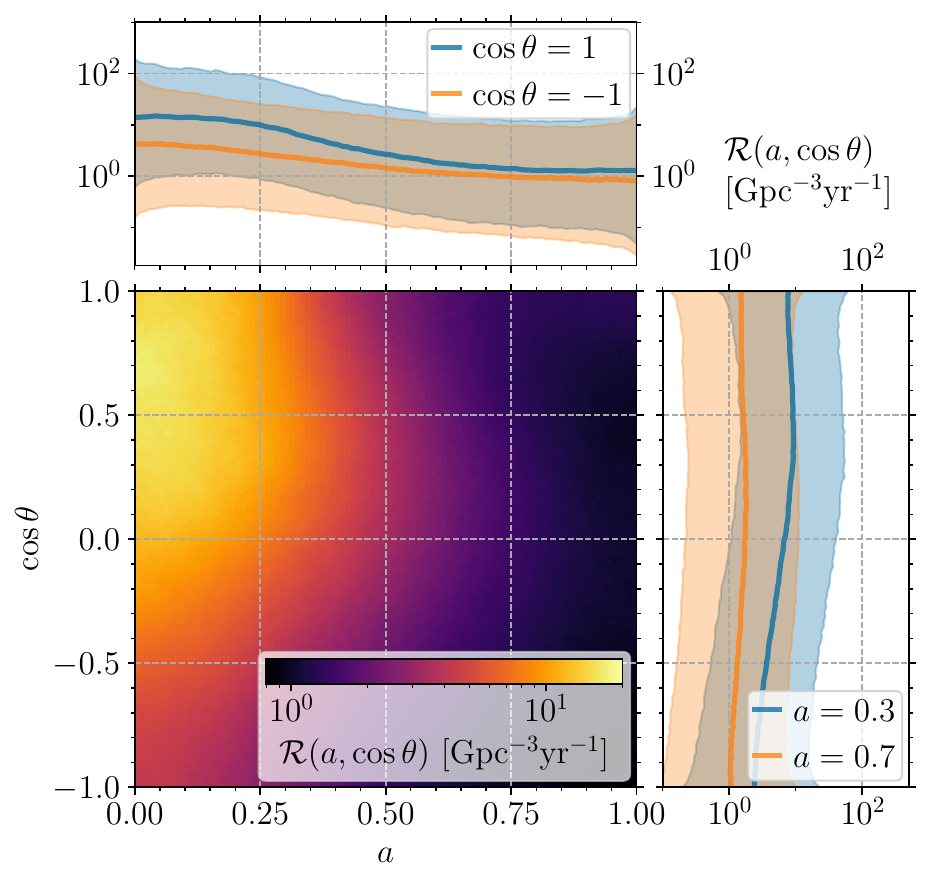}
\caption{Comoving merger rate density $\mathcal{R}(a,\cos\theta;z)$ as a function of BH spin magnitude $a$ and polar misalignment angle $\theta$ between the BH spin vector and binary orbital angular momentum, evaluated at a fixed redshift of $z=0.2$. The central panel displays the joint two-dimensional posterior median. The merger rate density evaluated at $\cos\theta=1$ (blue) and $\cos\theta=-1$ (orange) is shown as a function of $a$ in the upper panel, and evaluated at $a=0.3$ (blue) and $a=0.75$ (orange) as a function of $\cos\theta$ in the right-hand panel. The bold lines  the posterior medians while the shaded bands enclose the 90\% symmetric credible regions.}
\label{fig: at median}
\end{figure}

In Fig. \ref{fig: at median}, we show the inferred merger rate density over spin magnitudes and tilts evaluated at a fixed redshift $z=0.2$. There is a preference for more slowly spinning BHs positively aligned with the binary orbital angular momentum, as noted in our analyses of the
effective spin $\chi_\mathrm{eff}$ distribution. However, there are large uncertainties in the spin distributions, particularly the directions which are also consistent with being isotropic. Information about spins in the binary BH population mostly comes from effective spin degrees of freedom \cite{Miller:2024sui}, primarily $\chi_\mathrm{eff}$ which drives the parameter-estimation uncertainty and to a lesser extent the effective precessing spin parameter $\chi_\mathrm{p}$ \cite{Hannam:2013oca, Schmidt:2014iyl}. If there is no additional information in the data beyond the effective spins then any component-spin distributions that are consistent with the inferred effective-spin distributions are equally favored.

\vspace{0.1mm}

\begin{figure*}
\centering
\includegraphics[width=0.95\linewidth]{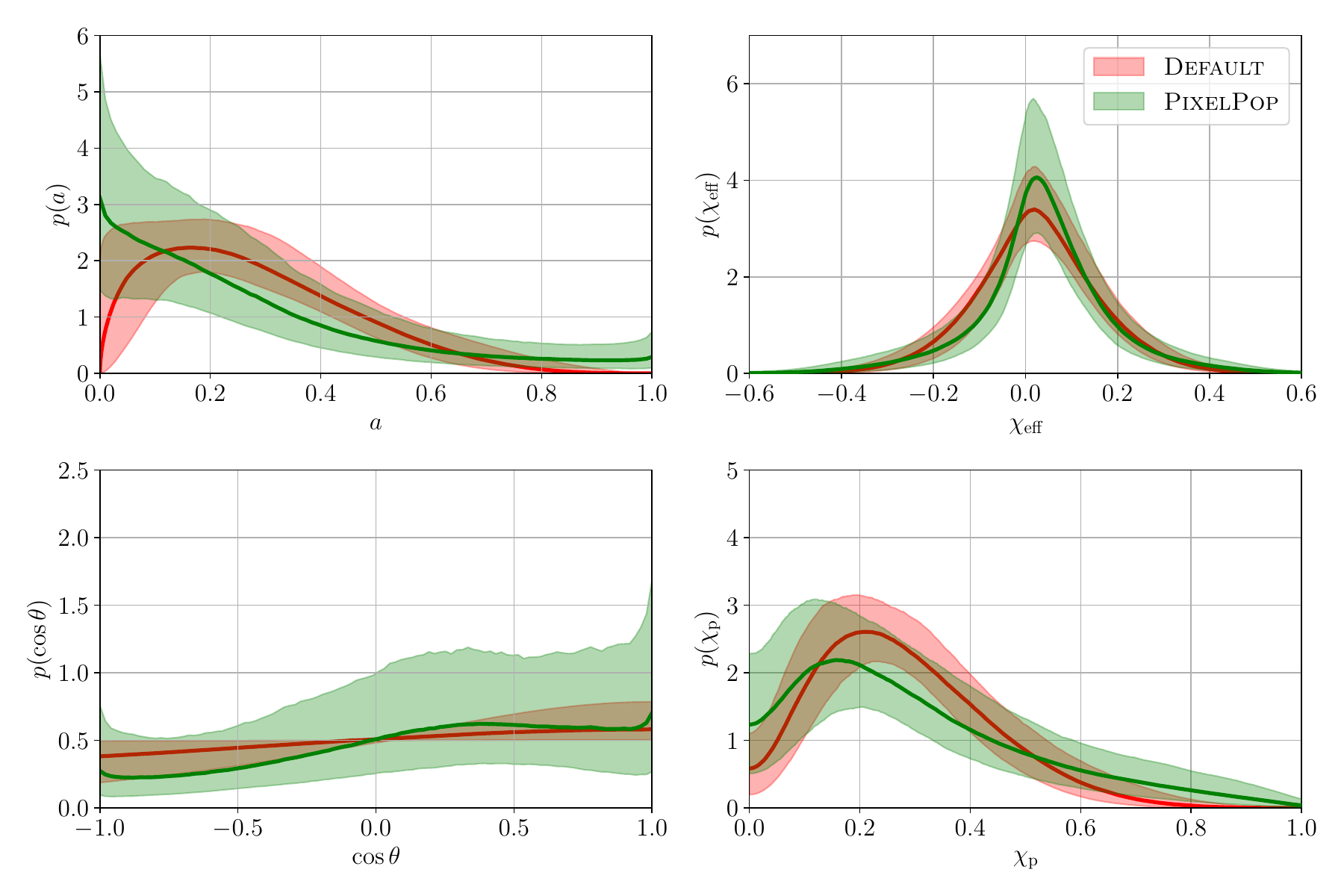}
\caption{Marginal distributions of component and effective BH spins inferred with GWTC-3. BH spin magnitudes $a$ and BH spin--orbit misalignment angles $\theta$ are shown in the top and bottom left panels, respectively. Effective aligned and precessing spins, $\chi_\mathrm{eff}$ and $\chi_\mathrm{p}$, are shown in the top and bottom right panels, respectively. The result of our nonparametric \textsc{PixelPop} analysis are shown in green and are compared to the parametric \textsc{Default} spin analysis of Ref.~\cite{KAGRA:2021duu}. The bold lines denote posterior medians while the shaded regions enclose the 90\% symmetric credible regions.}
\label{fig: spins}
\end{figure*}

We illustrate this in Fig.~\ref{fig: spins} by plotting the marginal spin magnitude and tilt probability densities inferred by \textsc{PixelPop} and the implied marginal distributions on the effective spins $\chi_\mathrm{eff}$ and $\chi_\mathrm{p}$, computed by renormalizing the merger rate density. These are compared to the \textsc{Default} spin parametric model from Ref.~\cite{KAGRA:2021duu}. Despite \textsc{PixelPop} having much larger posterior uncertainties compared to the parametric model, particular over $\cos\theta$, the uncertainties over $\chi_\mathrm{eff}$ are much smaller and comparable to the parametric analysis. Additionally, while the strong parametric $\textsc{Default}$ model enforces $p(a=0)=0$ a priori, with \textsc{PixelPop} we find that the spin magnitude distribution in fact peaks toward 0. Similarly, the $\cos\theta$ distribution is not required to peak at $\cos\theta=1$ \cite{Vitale:2022dpa}, in contrast to the assumptions of the $\textsc{Default}$ model. 

\end{document}